\documentclass[12pt,preprint]{aastex}
\pdfoutput=1
\received{2008}
\shorttitle{GCRT~J1742$-$3001}
\shortauthors{Hyman et al.}

\newcommand\src{\protect\objectname[]{GCRT~J1742$-$3001}}
\newcommand{\mjybm}{\mbox{mJy~beam${}^{-1}$}}

\begin{document}
\title{GCRT~J1742-3001: A New Radio Transient towards the Galactic Center}

\author{Scott D.~Hyman}
\affil{Department of Physics and Engineering, Sweet Briar College,
	Sweet Briar, VA  24595}
\email{shyman@sbc.edu}

%\author{Subhashis Roy}
%\affil{National Centre for Radio Astrophysics, Tata Institute of Fundamental Research, Pune, India}
%\email{roy@ncra.tifr.res.in}

\author{Rudy Wijnands}
\affil{Astronomical Institute 'Anton Pannekoek', University of Amsterdam, Kruislaan 403, 1098 SJ Amsterdam, The Netherlands}
\email{rudy@science.uva.nl}

\author{T.~Joseph~W.~Lazio}
\affil{Remote Sensing Division, Naval Research Laboratory, Washington, DC 20375-5351}
\email{Joseph.Lazio@nrl.navy.mil}

\author{Sabyasachi Pal}
\affil{National Centre for Radio Astrophysics, Tata Institute of Fundamental Research, Pune, India}
\email{spal@ncra.tifr.res.in}

\author{Rhaana Starling}
\affil{Department of Physics and Astronomy, University of Leicester, University Road, Leicester LE1 7RH, UK}
\email{rlcs1@star.le.ac.uk}

\author{Namir E.~Kassim}
\affil{Remote Sensing Division, Naval Research Laboratory, Washington, DC 20375-5351}
\email{Namir.Kassim@nrl.navy.mil}

\author{Paul S.~Ray}
\affil{Space Science Division, Naval Research Laboratory, Washington, DC 20375-5352}
\email{Paul.Ray@nrl.navy.mil}

\begin{abstract}
We report the detection of a new transient radio source, \src, located $\sim$1\arcdeg\
from the Galactic center. The source was detected ten times from late 2006 to 2007 May in
our 235 MHz transient monitoring program with the
Giant Metrewave Radio Telescope (GMRT). The radio emission
brightened in about one month, reaching a peak observed flux density of
$\sim$100 mJy on 2007 January 28, and decaying to $\sim$50 mJy by 2007 May when our
last monitoring 
observation was made. Two additional faint, isolated 235 MHz detections
were made in 
mid-2006, also with the GMRT. \src\ is unresolved at each epoch, with typical resolutions
of $\sim$20$\arcsec$$\times$10$\arcsec$. No 
polarization information is available from the observations. Based on nondetections in 
observations obtained simultaneously at 610 MHz, we deduce that the spectrum of \src\ 
is very steep, with a spectral index less than about -2.
Follow-up radio observations in 2007 September at 330 MHz and 1.4 GHz, and in 2008 February at 235
MHz yielded no detections.
No X-ray counterpart is detected in a serendipitous observation obtained
with the X-ray telescope aboard the {\it Swift} satellite during the peak of the
radio emission in early 2007. We consider the possibilities that \src\
is either a new member of an existing class of radio transients, 
or is representative of a new class having no associated X-ray emission. 

\end{abstract}


\section{Introduction}\label{sec:intro}

Many astronomical sources exhibit transient radio emission including flare stars, brown dwarfs, 
masers, gamma-ray bursts, pulsars, supernovae, active galactic nuclei, and neutron star and black hole
X-ray binaries. Unfortunately, there have only been a few blind radio surveys that have searched efficiently
for radio transients, due to one or more limitations including inadequate field of view, collecting area,
observing time, bandwidth, or time resolution.
Therefore, most radio transients have been found through monitoring observations of known or suspected transient emitters.
Recent improvements in low-frequency and wide-field imaging techniques, particularly at lower frequencies
\citep{lklh00, nlkhlbd04}, are enabling more
efficient searches for transient emission in the radio sky.    
Consistent with the expectation that previous limitations on the detection of radio transients were instrumental and
not astrophysical,
recent radio transient monitoring programs are revealing potentially new types of astronomical
sources including rotating radio transients \citep{mclaughlinetal06}, periodic, coherent bursts from an ultracool dwarf
\citep{hallinanetal07}, an extragalactic 30 Jy millisecond burst \citep{lorimeretal07}, 
several 1-3 Jy radio bursts reported at high and
low Galactic latitudes \citep{niinumaetal07, matsumuraetal07, kidaetal08}, and ten milliJansky transients detected in 22 years of archival VLA observations of a single field of view
at 5 and 8.4 GHz \citep{boweretal07}.

Over the past several years, we have conducted a blind search for radio transients in the Galactic
center (GC) at 235 and 330 MHz using archival VLA observations made between 1989 and the present,
and monitoring observations with both the VLA and GMRT beginning in 2002.
Our motivation is that the naturally wide fields of view obtained at lower frequencies combined with the high stellar densities
toward the GC provide an efficient means for searching for radio transients.
We detected two radio transients, GCRT J1746-2757 \citep{hlkb02} and
GCRT J1745-3009 \citep{hlkrmy-z05, hlrrkn06, hrplrkb07, kaplanetal08}, for which no counterpart has been detected at high energies,
and therefore would not have
been detected in the more conventional manner of following up newly discovered X-ray or $\gamma$-ray transients.  
These two radio transients have markedly different observed properties. GCRT J1746-2757 was detected in a single 330 MHz, 7-hr
observation in 1998
with a constant $\sim$200 mJy flux density, while GCRT J1745-3009 emitted five $\sim$1 Jy, $\sim$10 minute-long bursts 
every 77 minutes in a 330 MHz observation in 2002, and a single, weaker burst in each of two subsequent observations in 2003 and 2004.
The properties of GCRT J1745-3009, discussed in detail in \cite{hrplrkb07}, suggest strongly that unlike most
radio transients which are incoherent synchrotron emitters, GCRT J1745-3009 is a member of a new class of coherent
emitters.

This paper reports the discovery of another low frequency radio transient, \src, detected at 235 MHz in our 2006-2007
monitoring program with the Giant Metrewave Radio Telescope (GMRT). The source is not detected in
X-ray observations obtained with the X-ray telescope
(XRT) aboard {\it Swift} taken in early 2007 during the peak of the radio emission. Its temporal evolution is similar to that for the
bright radio transient, the Galactic Center Transient (GCT), which also has not been detected at high energies \citep{zhaoetal92}. 
We present the observations and results on \src\ in Section 2, and discuss similarities to the GCT and constraints on possible models
in Section 3.

\section{Observations and Results}\label{sec:observe}

\subsection{Radio Detection}

Our 2006-2007 GMRT monitoring program of the Galactic center consists of twenty-one
simultaneous 235 and 610 MHz observations, each 2 to 5~hr long, for a total observing time
on source of 66~hr.
The observations were pointed towards the bursting Galactic
center radio transient, GCRT J1745-3009, in order to monitor it for renewed activity (none was detected) while at
the same time searching for serendipitous new transients enabled by the large GMRT field-of-view at 235 MHz (2.0\arcdeg\ FWHM).

Initial calibration and imaging was performed in a manner
similar to that described in \cite{hlrrkn06}.
Amplitude calibration was conducted in reference to
3C286 and phase calibration was based on observations of the nearby source 
J1830-360. A polyhedral imaging algorithm was employed to compensate for the non-coplanarity
of the GMRT at low frequencies \citep{cp92} and several iterations of imaging,
deconvolution, and self-calibration were used.

\begin{figure}
\begin{center}
\epsscale{0.45}
\includegraphics[height=3.5in]{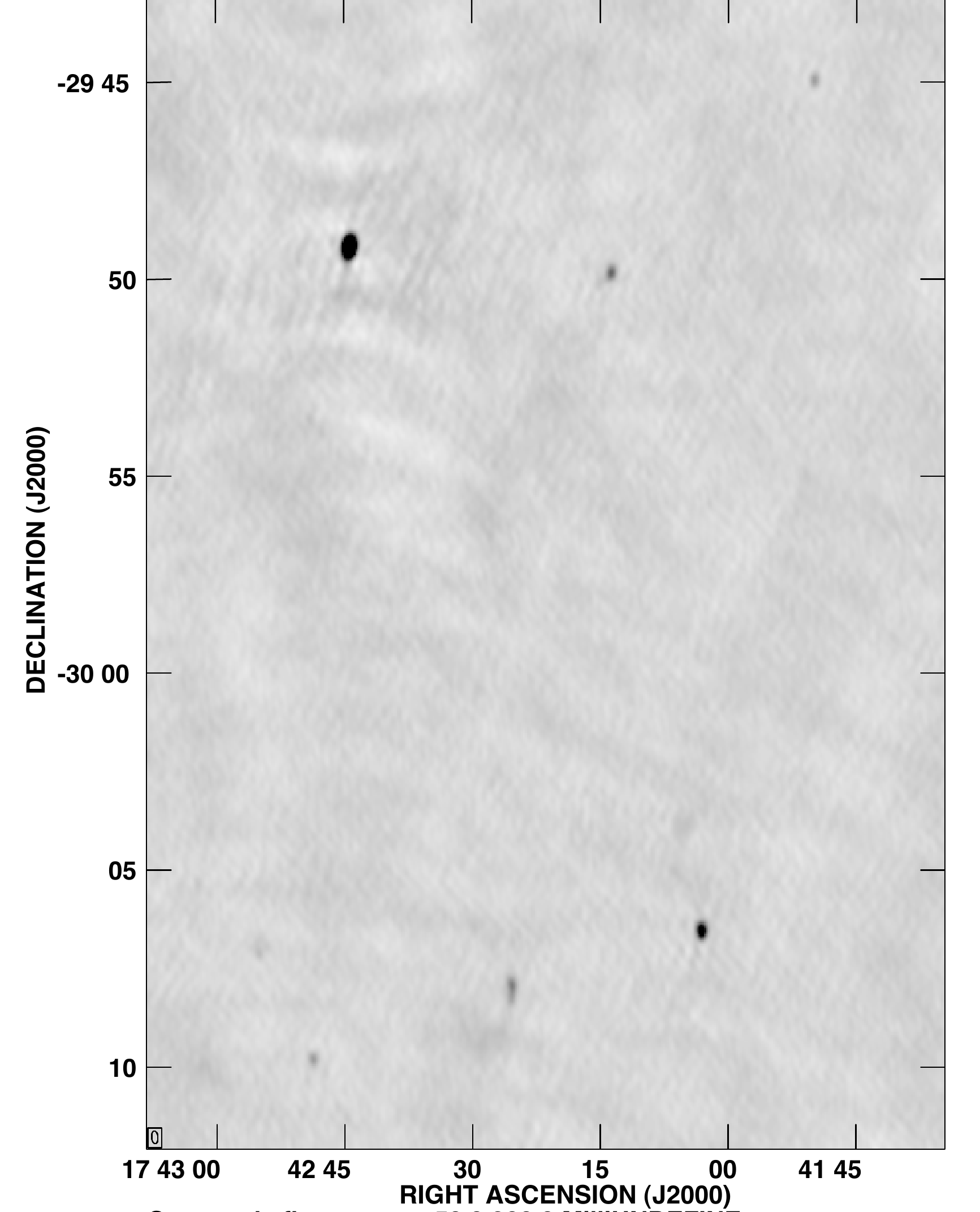}
\includegraphics[height=3.5in]{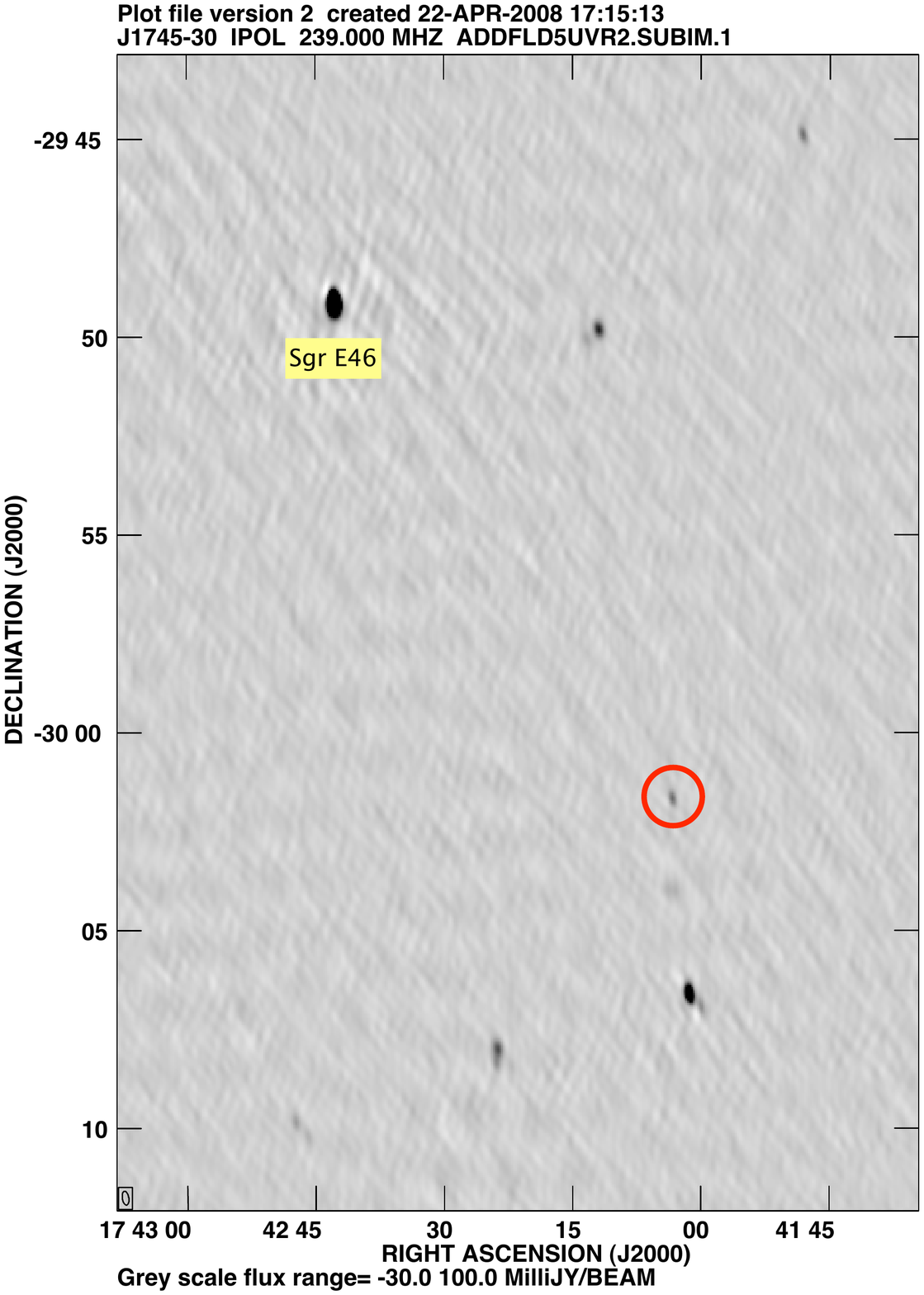}
%\plotone{08MAR2006NOTRANS_crop.pdf}
%\plotone{28FEB2007TRANS_crop.pdf}
%\rotatebox{-90}{\plotone{f2a.ps}}\,
%\rotatebox{-90}{\plotone{f2b.ps}}\\
%\rotatebox{-90}{\plotone{f2c.ps}}
\end{center}
\caption[]{235 MHz GMRT image of the region surrounding \src\ on 8 March 2006 before detection
 (\textit{left}) and on 2007 January 28 during the peak of the
 transient emission (\textit{right}).  For the 2006 March 8 image, the
 rms noise level is 4.6 mJy~beam${}^{-1}$ with a resolution of
 $17\farcs6 \times 9\farcs4$; for the 2007 January 28 image, the rms
 noise level is 6.3 mJy~beam${}^{-1}$ with a resolution of $21\farcs1 \times 9\farcs9$.
 Both observations are approximately 3~hr in duration.  For reference,
 the strong source Sgr~E46, used to determine the relative flux density correction for
 \src\ at each epoch (see Section 2.1), is labeled in the 2007 January 28 image. 
 The bright source,
 Sgr~E18, located $\sim$4\arcmin\ south of \src\ was used for astrometric corrections
 and to double check the relative flux density corrections. The faint source, Sgr~E19,
 located halfway between \src\ and Sgr~E18 is discussed at the beginning of Section 3.}
\label{fig:onoffimage}
\end{figure}

\begin{figure}
\epsscale{1.25}
\includegraphics[width=6.0in]{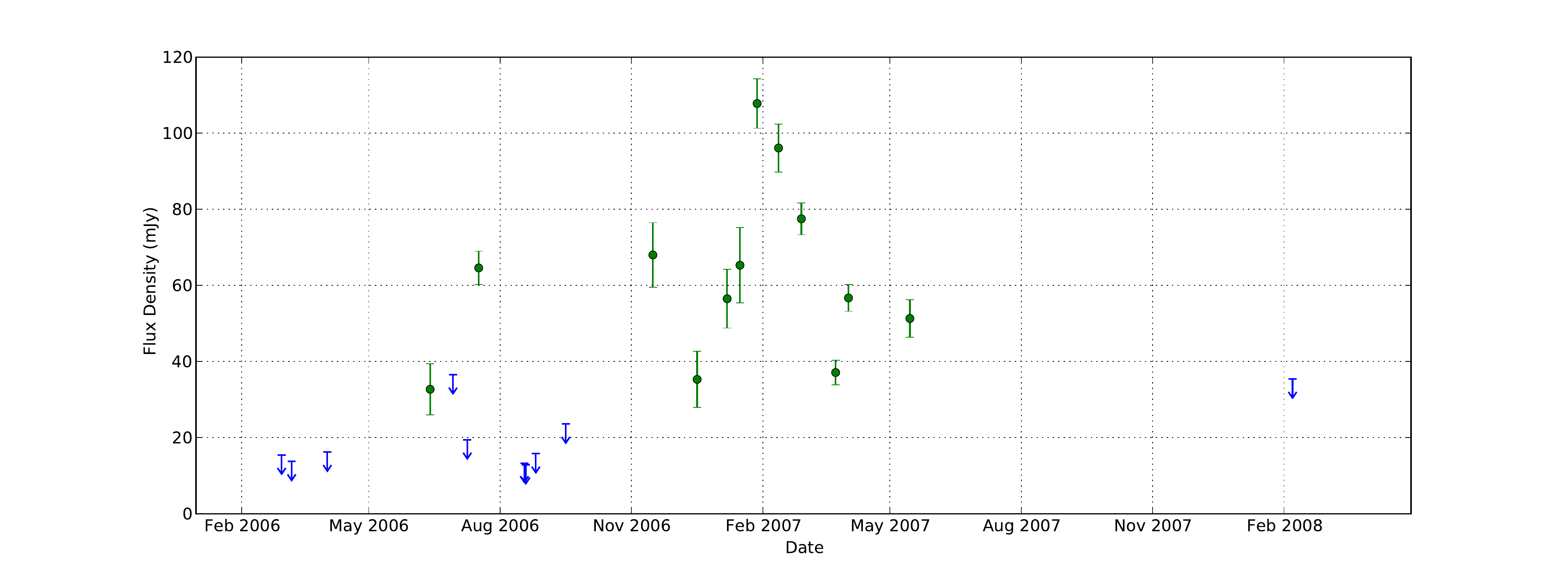}
\caption{The 235 MHz light curve of the GMRT detections (circles) of \src\ and 3$\sigma$
upper limits.} 
\label{fig:lightcurve}
\end{figure}

The new transient radio source, \src, reported in this paper is derived from the 235 MHz GMRT monitoring
observations, which are listed in Table~\ref{tab:log}. Figure~\ref{fig:onoffimage} shows \src\ located 0.1$\arcdeg$ above the
Galactic plane and 1.25$\arcdeg$ southwest of Sgr A*, toward the Sgr E complex of compact sources.
Figure~\ref{fig:lightcurve} shows the light curve of \src, including 3$\sigma$ upper limits for nondetections.
The source exhibited potentially multiple short bursts (epochs 2006 June 13, 2006 July 17, and possibly 2006 November 16), of a duration
no more than about 10 days, followed by a long burst
beginning on 2007 December 17 and continuing until at least 2007 May 15 when our 2007 monitoring campaign ended.  This
last burst appears to have a fast rise, reaching the peak in about
1 month, and then decaying over the next 4 months.
The peak measured flux density ($107.9 \pm 8.4$ mJy) occurred on 2007 January 28. 
No polarization information is available from the observations.
A correction of 1.37$\times$ for primary beam attenuation at the location of \src\ has
been applied to the flux densities, and additional corrections are discussed below. 

\src\ is unresolved at a nominal resolution of $\sim$20$\arcsec$$\times$10$\arcsec$ obtained at each epoch.
Fitting an elliptical Gaussian, along with a background level, to the source yields an average position and 1$\sigma$ uncertainty 
of (J2000) $17^{\mathrm{h}}$ $42^{\mathrm{m}}$ 4\fs67 ($\pm 0\fs57$), 
$-30\arcdeg$ 01\arcmin\ 44.5\arcsec\ ($\pm 2.5\arcsec$). This position includes corrections for ionospheric
refraction which is prevalent at low radio frequencies. To determine these corrections, we first measured the position at each epoch of 
the bright extragalactic source, Sgr E18 \citep{grayetal93, crametal96}, located only $\sim$5\arcmin\ south of \src\ and visible on
Figure~\ref{fig:onoffimage}. We then subtracted the average of these positions
from the average of the 6 and 20 cm positions of this source determined in the reanalysis by \cite{whiteetal05} of 
their Galactic plane radio survey of 586 compact sources \citep{zoonetal90, helfandetal92, beckeretal94}.   
The resulting differences of +0\fs71 $\pm 0\fs43$ in right ascension and -3.2\arcsec\ $\pm 1.6\arcsec$ in declination were applied
to the average position obtained for \src, yielding the corrected position given above. As a check,
we also followed the same procedure for the bright source, Sgr E46, located further from \src\ as
indicated on Figure~\ref{fig:onoffimage}. In principle, ionospheric refraction could be position
dependent over the large field of view at 235 MHz, but we find the position correction for Sgr E46 to be consistent with that for Sgr E18.

Based on our experience with the bursting transient GCRT~J1745$-$3009 \citep{hlkrmy-z05}, we also searched for
flux density variations within the scans comprising each day's
observation of \src.  A typical observation consisted of numerous scans, each
approximately 30 minutes in duration.  On 2007
February 28, the flux density measured for
the third scan is $107 \pm 12$ mJy, a $3.3\sigma$ marginal variation in comparison to the 68 mJy average of the other six scans. 
No significant variation is found on shorter timescales within the third scan, and no other 
scan-to-scan marginal variations are detected at this or any of the other epochs. 

The error bars in Figure~\ref{fig:lightcurve} reflect the rms noise levels ($\sim$3-10 \mjybm) of the images near the location of \src\
and a 5\% uncertainty in the absolute calibration of each data set added in quadrature. 
The latter was determined by comparing, from epoch to epoch, the flux densities of
known bright sources located near \src\ that should be constant in time. In the process, we discovered that the flux densities of
these field sources varied 
significantly, but approximately in unison, due to an instrumental effect (the increase in
system temperature from the calibrator field to the target) which has been encountered by
other GMRT observers (e.g., \cite{royrao04}).
Following the approach of \cite{vanderhorst08}, we therefore applied a relative correction
to \src\ tied to the flux density of the relatively bright extragalactic source
Sgr E46 (indicated on Figure~\ref{fig:onoffimage}) determined for each epoch and
at the peak of the emission from \src\ on 2007 January 28.
The correction factors range from 0.4 to 2.1 and are provided in Table~\ref{tab:log}. As a check, we 
also corrected the flux densities of the nearby source Sgr E18, mentioned above, on which we based the position 
correction. 
The residual epoch to epoch variations are at the $\sim$5\% level for this source,
and we therefore include this relative calibration correction uncertainty in the error bars of Figure~\ref{fig:lightcurve}. 

To determine an overall calibration correction to apply to the light curve of \src, 
we used a separate 235 MHz observation provided by S.~Roy from a separate program \citep{royrao06},
but which does not suffer
from absolute calibration limitations.
The latter observation was pointed 1.2\arcdeg\ north of the monitoring observations, and so the region containing \src\ and
the Sgr E complex lies outside the FWHM of the primary beam.
We chose a bright source (TXS 1745-296) located at the half-power point of both primary beams as
the most suitable source on which to base the overall correction.   
The ratio of the flux density of TXS 1745-296 (0.6 Jy) to that measured from our observation of
28 January 2007 was determined to be 1.43,
and was used to further correct the flux densities of \src. Both the relative epoch to epoch and overall corrections are
reflected in the light curve
of Figure 2. The error bars, however, do not include an additional $\sim$5\% uncertainty in the overall correction of the light curve in order to not mask the relative changes from epoch to epoch.

We searched for \src\
in our simultaneous GMRT observations acquired at 610 MHz, but were unable to detect it at any epoch.
The location of
\src\ is 40.0\arcmin\ from the phase center of the observations; this is far outside the
44\arcmin\ field of view (FWHM) at 610 MHz where the primary beam attenuation is much
greater than the attenuation factor of~1.4 at~235~MHz.
Models of the primary beam pattern do not provide reliable
correction values at such extreme distances from the phase center.
However, in principle, a correction factor can be estimated, and hence, also the upper
limit on the flux density of \src, by using the ratio of the measured to the actual
flux density of a field source located sufficiently near to \src. While we were 
unable to obtain any such known 610 MHz flux densities, we were able to estimate the 610 MHz
flux density of the nearby source, Sgr E18, from a power law fit to the flux 
densities available at 330, 1281, 1658, and 4850 MHz \citep{nlkhlbd04, lc98},
together with the average flux density determined from our 235 MHz observations
(which varied by only $\sim$5\%, as discussed above).
Sgr E18 is detected in our 610 MHz observations
at the 20$\sigma$ level, and is fortuitously 
located only $\sim$5\arcmin\ south of \src\ and at nearly the same angular distance (39.6\arcmin)
from the phase center of the observations.  This close proximity mitigates against the inaccuracies associated with the
asymmetries and high radial dependence of
the primary beam correction known to exist far outside the nominal field of view; hence, we could apply the correction 
obtained for Sgr E18 to the noise level of the image in order to estimate an upper limit for the 610 MHz flux density of \src.

The power law fit
($S \propto \nu^{\alpha}$) to the flux densities of Sgr E18 yields a spectral index of $\alpha = -0.87 \pm 0.08$, and a predicted flux density of $225 \pm 20$ mJy at 610 MHz. The 
measured flux density of Sgr E18 at 610 MHz from the 28 January 2007 epoch 
(the date of the peak emission
at 235 MHz) is $14.5 \pm 0.8$ mJy, 
leading to a correction factor of $15.5 \pm 1.6$. (Note that this factor accounts for the primary beam
attenuation and also any absolute
calibration error that might exist in the 610 MHz data analogous to that
discussed above for the 235 MHz observations.)
We applied this correction to the 0.31~\mjybm\ rms
noise level of the uncorrected image. The 610 MHz 3$\sigma$ upper limit for \src\ is therefore
$\sim$15 mJy on 2007 January 28, compared to the 107 mJy detection at 235 MHz obtained
for that epoch. This result yields a spectral index constraint of $\alpha \lesssim -2$. 
We also attempted to determine the spectrum of \src\ by looking for any
change across the 6-MHz bandpass of the 235~MHz observations. A power law fit
results in a largely unconstrained spectral index of $\alpha = 0 \pm 5$.

\src\ is not detected in follow-up observations at 330 MHz (VLA) and 1.4 GHz (GMRT) in 2007 September, with 3$\sigma$ upper
limits of 30 mJy and 0.9 mJy, respectively. If the source was still active during these observations with a 235 MHz flux density
of $\sim$50 mJy, then the spectral index of the source at that time would have been steeper than $\alpha = -2$. 
However, \src\ could also have faded significantly from when it was last detected in 
2007 May. 
Unfortunately, we could not obtain follow-up GMRT observations at 235 MHz until
2008 February 7 when the source was not detected, with a 3$\sigma$ upper limit of 36 mJy. 

In addition to our 235 MHz GMRT monitoring program, we also monitored the Galactic center with the VLA at 330 MHz during 2006, but not 2007. 
The VLA program, which will be presented in detail in a subsequent paper,
consisted of twenty observations from 2006 February to 2006 September, each
$\sim$2.5~hr in duration, for a total of $\sim$50~hr. \src\ was not detected in any of the epochs,
although one of the 2006
observations occurred on 2006 June 13, the same date as when the GMRT detected the transient
with a 33 mJy
flux density at 235 MHz. Unfortunately, the VLA $3\sigma$ upper limit of 45 mJy barely constrains
the spectral index between 235 and 330 MHz
($\alpha < +0.9$). However, following the method described above, we estimate an upper limit of $\sim$10 mJy for the 610 MHz flux
density for the 2006 June 13 epoch, which, given the 235 MHz value, corresponds to a spectral index constraint of $\alpha \lesssim -1.3$.

An image made from the
combination of our 2006 VLA 330 MHz observations did not 
detect the source with a $3\sigma$ upper limit of 10 mJy. The upper limit obtained by combining our 235 MHz
nondetection observations from 2006 is also 10 mJy.

\subsection{X-ray Nondetection}

On 2007 February 2, the {\it Swift} satellite serendipitously
pointed with the XRT for $\sim$190 seconds towards the region of the radio transient.
No X-ray
source was detected in the field-of-view of the XRT. Using the
Bayesian confidence limit method \citep{kbn91} we calculated a 3$\sigma$
upper limit on the count rate of
0.038 counts/s (for the energy range 0.3-10 keV) at the
position of the radio transient.

To convert this count rate upper limit to a flux upper limit we
used WebPIMMS\footnote{http://heasarc.gsfc.nasa.gov/Tools/w3pimms.html}
and we assumed an absorbed power law spectrum with
a column density\footnote{http://heasarc.gsfc.nasa.gov/cgi-bin/Tools/w3nh/w3nh.pl}
$N_{h} = 1.3 \times 10^{22}$ cm$^{-2}$ and a photon index of 2. This resulted in upper limits on the 2-10 keV
absorbed flux of $2.3 \times 10^{-12}$ erg cm$^{-2}$ s$^{-1}$ and 
an unabsorbed flux of
$2.5 \times 10^{-12}$ erg cm$^{-2}$ s$^{-1}$.
For a harder spectrum (photon index $\sim$1) these
values are around 70\% larger. If the source is located near the
Galactic center at a distance of $\sim$8.5 kpc, then the upper limit
on the 2-10 keV luminosity would be $\sim 2.2 \times 10^{34}$ erg s$^{-1}$,
assuming a photon index of 2.

In addition to the 2007 February 2 observation, {\it Swift} was pointed
at the source direction on four additional occasions from 2005 to 2008 listed in
Table~\ref{tab:xraylog}.
We combined all five data sets but still the source was not detected.
The 3$\sigma$ upper limit for the combined data set is 0.010 counts/s.
Assuming the same spectral shape as above, the absorbed flux
upper limit (2-10 keV) would then be $5.9 \times 10^{-13}$ erg cm$^{-2}$ s$^{-1}$ 
and the unabsorbed flux limit would be $6.7 \times 10^{-13}$ erg cm$^{-2}$ s$^{-1}$. For
a distance of 8.5 kpc, the 2-10 keV luminosity upper limit would be $5.5 \times 10^{33}$
erg s$^{-1}$.

\section{Discussion}\label{sec:Discussion}

We have searched the environment of \src\ for associated discrete soures or extended structures.
\src\ is located in the Sgr E complex of discrete sources, many of which have a flat spectrum
and are confirmed by recombination line observations to be HII regions \citep{crametal96,
grayetal93}. The closest source is Sgr E19, which is faintly visible about 2\arcmin\ south
of \src\ on Figure~\ref{fig:onoffimage}. This source is described
as a possible candidate young supernova remnant in \cite{crametal96}. However, more extensive observations and the
detection of counterpart infrared sources \citep{misanovic02} indicate that Sgr E19 is
much more likely to be a HII region and that it is not associated with Sgr E, but rather is much closer to us. Whether \src\
is within the Sgr E complex or is also much closer remains to be seen. We see no other nebulosity on radio images of
this region that appears to be connected to \src, although there are several 2MASS infrared sources located within the $3\sigma$
positional error ellipse, the nearest 10\arcsec\ away from \src.

We have used the upper limit to the angular size of \src\ to constrain its brightness temperature.
Additional Gaussian fits were made to the
2007 January 28 detection, with the major and minor axes varied  
until the fit returned an integrated flux density $1\sigma$ above the nominal integrated flux.
The corresponding upper limit on the geometric mean of the deconvolved major and minor axes (FWHM)
is 8\arcsec. In turn, this implies a lower limit on its brightness
temperature of $1.2 \times 10^{4} K$. Since this is a conservative estimate, we conclude that \src\ is
likely nonthermal, which is also consistent with the steep spectrum determined in Section 2.1.

\subsection{Similarity to the Galactic Center Transient}

\begin{figure}
\epsscale{1.25}
\includegraphics[width=6.0in]{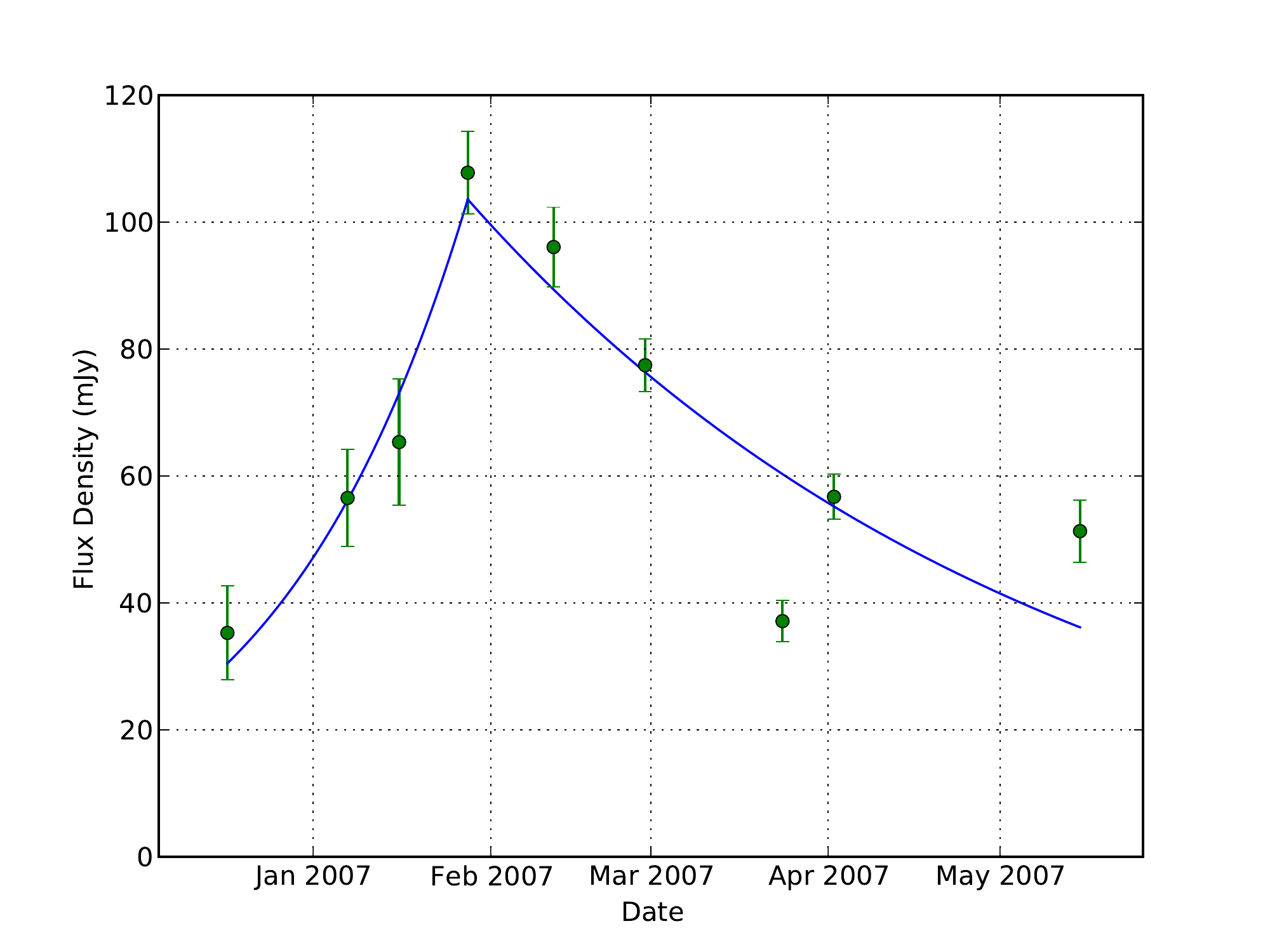}
\caption{Exponential fits to the 235 MHz light curve of \src.
The rise and decay time constants resulting from the fit are
$34 \pm 10$ days and $102 \pm 38$,
respectively, and the peak flux density is 28 January 2007 $\pm$ 5 days.}
\label{fig:lcfit}
\end{figure}

The temporal evolution of \src\ is similar to that for the GCT
which was detected in monitoring observations of Sgr A* from late 1990 December until late 1991 September at radio wavelengths from 1.3 to 22 cm \citep{zhaoetal92}.
The GCT reached its maximum in approximately one month and faded with a time scale of about three months. 
In Figure~\ref{fig:lcfit} we show exponential fits to the rising and decaying portions of
the light curve of \src. The rise and decay time constants resulting from the fits are
$34 \pm 10$ days and $102 \pm 38$ days,
respectively, and the date of the peak flux density determined is 2007 January 28 $\pm$ 5 days.
A power law $S \propto t^{-\beta}$ fit to the six detections from 2007 January 28 through 
2007 May 15 yields
an index of
$\beta = 0.60 \pm 0.14$, similar to that obtained ($0.67 \pm 0.08$) for the decay portion of the
light curve of the GCT.

The GCT had a $\sim$1 Jy peak flux density in the wavelength range 18-22 cm ($\sim$1.5 GHz), and a relatively constant and 
steep spectral index of $\alpha = -1.2$ over the duration of the transient emission. 
\src\ has a $\sim$100 mJy peak at 235 MHz and also has a steep spectrum with $\alpha \lesssim -2$ 
calculated between 235 and 610 MHz. Assuming the spectral index is constant, the corresponding 
1.5 GHz upper limit is $\sim$2.5 mJy, or about 
400$\times$ fainter than the GCT. However, the distance to \src\ is unknown, whereas
\cite{zhaoetal92} determined that the GCT is located
at the Galactic center, about 8.5 kpc distant; if \src\ is located much closer
to us than the GC, its luminosity could be even smaller compared to the GCT.

Extrapolating the 1.5 GHz flux density upper limit for \src\ with either the exponential or power law fits yields flux densities
which are consistent
with our 1.5 GHz nondetection on 2007 September 19, described earlier.
We note, though, that the GCT exhibited a significant secondary maximum in the 
18-22 cm observations about 6 months after the primary. If \src\ also emitted a secondary maximum and it
occurred during the second half of 2007 when we were no longer monitoring, our sparse follow-up radio
observations in 2007 September and 2008 February might very well have missed this additional activity.
Similarly, it is possible that the weaker flux densities recorded on 2007 March 24 and
2006 December 17
(see Figure~\ref{fig:lightcurve}) are indicative of broad
secondary maxima extending before and/or after our observations on 2006 November 16 and
2007 May 15,
respectively. However, it is also possible that these low data points are just  
short fluctuations in the light curve,
as was also seen in the light curve of the GCT up to the 50\% level. Importantly, the two
detections in mid-2006 demonstrate the recurrence of transient emission from \src, although it is
not clear how and/or whether these fainter detections are related to the primary emission in 2007.

Based on its radio and infrared properties, \cite{zhaoetal92} suggest
that the GCT is a synchrotron-emitting radio transient associated with an X-ray binary system.
Indeed, the duration of the radio outburst from the GCT (and from \src) is similar to those
observed for X-ray outbursts from accreting black hole systems.
But while the {\it Swift}
X-ray nondetection of \src\ significantly constrains the level of any X-ray emission from it,
there are unfortunately no reports in the literature of X-ray emission upper limits for the GCT. 
Without such information, we cannot determine if the temporal similarities of \src\ and the GCT are
merely coincidental, or if the two transients are actually similar in nature, in which case 
the interpretation of the GCT as an X-ray binary system might need to be reexamined.

\subsection{X-ray Model Constraints}

For the majority of black hole X-ray binaries
a 'universal correlation' has been seen 
between the radio and X-ray luminosities of the systems 
\citep{corbel03, gallo03}. Assuming the relation found by \cite{gallo03} for much
higher frequencies (4.9 to 15 GHz) is valid at 235 MHz, our $\sim$100 mJy peak flux density
corresponds to a predicted 2-11 keV X-ray flux of
$(0.5 - 2) \times 10^{-8}$ erg cm$^{-2}$ s$^{-1}$ for \src\ if it is located at 1 kpc, or
$(0.3 - 1) \times 10^{-7}$ erg cm$^{-2}$ s$^{-1}$ if the
distance towards the source is 8.5 kpc. This is $\sim$4 orders of magnitude
brighter than the X-ray upper limit we deduced in Section 2.2 for \src. Although a number of outliers have
been identified which do not follow this correlation, they are always under
luminous in the radio compared to their X-ray luminosity and not under luminous in the
X-ray.

We note, however, that black hole X-ray binaries in low luminosity states typically have flat radio spectra
\citep{fender01}, which we do not find is the case for \src. 
Instead, and as described in Section 2.1, we find the source apparently has a very steep 
radio spectrum towards lower frequencies. The GHz flux density is therefore probably much 
lower than that observed at 235 MHz, which in turn would significantly lower the 
inferred X-ray flux. Indeed, taking the upper limit of the spectral index of \src\ ($\alpha$ = -2) to extrapolate
the flux density to 4.9 GHz, yields a corresponding X-ray flux upper limit approximately equal to the upper limit of the  
{\it Swift} observation. Therefore, it is possible that \src\ is a black hole binary and that, instead of being in a 
low luminosity state, the source did emit a faint, undetected X-ray outburst during our radio detection. We underscore, though,
that our constraint on the radio spectrum of \src\ is determined only from the 235 MHz detection and 610 MHz nondetection. 
If the spectral index is not uniform, then depending on the extent to which it flattens toward higher frequencies,
we might have been able to detect an X-ray outburst.

Alternatively, the low-frequency radio
emission could have had a significant delay with respect to a possible
undetected bright X-ray outburst.
However, in that case the radio
spectrum should also steepen toward lower radio frequency since if
this was not the case the associated X-ray flux would 
be very high.
It is unlikely that such a bright X-ray outburst would have
gone unnoticed by the X-ray monitoring instruments in orbit especially
if the X-ray outburst had a similar duration as the radio outburst.
Perhaps the simplest explanation is to assume that the new radio transient is not an accreting black hole at all,
although it might also be one of the first of a group of accreting black holes which
are radio bright and very under-luminous in the X-ray. 

Alternatively, the new radio transient might be an accreting neutron star,
but since such systems are fainter in the radio with respect to
their X-ray fluxes compared with the accreting black holes, the discrepancy
between the expected and observed X-ray fluxes would even be larger
than for the black holes.

In summary, we have detected a new low frequency radio transient, \src, with the GMRT which is the third X-ray quiet transient found in our
Galactic center radio monitoring program. Each of these transients exhibited markedly different observational properties: 
(1) single epoch detection at 330 MHz (GCRT J1746-2757); (2) repeated, and possibly coherent, bursts over a six hour
330 MHz observation and fainter single bursts
within the subsequent two years (GCRT J1742-3009); and (3) multiple detections at 235 MHz, including one possible 25 min. coherent burst,
brightening and fading over a period of six months, with a steep spectrum (\src).

Continued radio monitoring of these transients and multi-wavelength follow-up observations are
necessary to identify their emission mechanisms and to determine whether they are examples of new 
populations of astronomical objects or are related in a hitherto unknown way to existing source classes.

\acknowledgements

We thank the staff of the GMRT that made these observations possible.
GMRT is run by the National Centre for Radio Astrophysics of the Tata
Institute of Fundamental Research. S.D.H.\ is supported by funding from
Research Corporation and SAO Chandra grants GO67135F and GO67038B. Basic research in radio
astronomy at the NRL is supported by 6.1 base funding. R.S. acknowledges support from the Science and 
Technology Facilities Council.

\begin{deluxetable}{lcccc}
\tablecaption{235 MHz GMRT Observations of \src\  \label{tab:log}}
\tablewidth{0pc}
\tabletypesize{\small}
\tablehead{
\colhead{Epoch}
	& \colhead{Duration} 
	& \colhead{Rel. Calib.}
	& \colhead{Resolution}
	& \colhead{Flux Density\tablenotemark{a}} \\
%	& \colhead{5$\sigma$ Upper Limit} \\

	& \colhead{(hr)}
	& \colhead{Factor}
	& \colhead{($\arcsec$)}
	& \colhead{(mJy)}}
%	& \colhead{(\mjybm)}}
	
\startdata
2006 March~01 & 3.0 & 0.99 & $15.5 \times 10.2$ & $5.1$\tablenotemark{b} \\
2006 March~08 & 3.3 & 2.12 & $17.6 \times 9.4$ & $4.6$\tablenotemark{b} \\
2006 April~2-3 & 3.5 & 0.64 & $15.5 \times 10.0$ & $5.4$\tablenotemark{b} \\
2006 June~13 & 4.0 & 0.53 & $26.9 \times 9.4$ & $32.7 \pm 6.7$ \\
2006 June~29 & 2.4 & 2.14 & $20.4 \times 11.0$ & $12.2$\tablenotemark{b} \\
2006 July~09 & 2.6 & 0.78 & $18.6 \times 10.2$ & $6.5$\tablenotemark{b} \\
2006 July~17 & 3.0 & 0.83 & $17.2 \times 9.3$ & $64.6 \pm 4.4$ \\
2006 August~18 & 3.5 & 0.53 & $31.6 \times 12.5$ & $4.4$\tablenotemark{b} \\
2006 August~19 & 5.6 & 0.81 & $20.5 \times 9.2$ & $4.3$\tablenotemark{b} \\
2006 August~26 & 2.7 & 0.76 & $22.8 \times 9.3$ & $5.3$\tablenotemark{b} \\
2006 September~16 & 4.2 & 1.64 & $18.9 \times 12.0$ &  $7.9$\tablenotemark{b} \\
2006 November~16 & 2.3 & 0.82 & $24.2 \times 11.4$ & $68.0 \pm 8.5$ \\
2006 December~17 & 3.1 & 0.47 & $24.6 \times 8.7$ & $35.3 \pm 7.4$ \\
2007 January~07 & 2.5 & 0.39 & $27.3 \times 9.1$ &  $56.5 \pm 7.7$ \\
2007 January~16 & 2.6 & 0.53 & $21.4 \times 8.9$ & $65.3 \pm 9.9$ \\
2007 January~28 & 2.9 & 1.00  & $21.1 \times 9.9$ & $107.8 \pm 6.5$     \\
2007 February~12 & 2.5 & 1.07 & $17.3 \times 9.3$ & $96.1 \pm 6.3$ \\
2007 February~28 & 3.0 & 0.58 & $14.9 \times 10.0$ & $77.5 \pm 4.2$ \\
2007 March~24 & 2.8 & 0.75 & $16.4 \times 9.3$ & $37.1 \pm 3.2$ \\
2007 April~2-3 & 2.7 & 0.90 & $17.5 \times 10.4$ &  $56.7 \pm 3.5$ \\
2007 May~15 & 3.9 & 0.96 & $16.5 \times 10.7$ & $51.3 \pm 4.9$ \\
2008 February~07 & 3.2 & 0.55 & $28.6 \times 9.0$  & $11.8$\tablenotemark{b} \\ 
\enddata	
\tablenotetext{a}{The flux densities include an overall calibration correction of 1.43$\times$ in
addition to the relative calibration corrections (see text).}
\tablenotetext{b}{No detection. The rms noise level measured at the location of \src\ is listed.}

\end{deluxetable}

\clearpage

\begin{table}
\caption{Upper limits on the 0.3-10 keV X-ray count rates and 2-10 keV
X-ray fluxes during five serendipitous {\it Swift} XRT observations. Fluxes are derived assuming an absorbed power
law spectrum with $\Gamma=2.0$ and $N_{h}=1.3\times 10^{22}$
cm$^{-2}$.}
\begin{center}
\begin{tabular}{lcccc} \hline
Epoch & T$_{\rm exp}$ & count rate & absorbed flux & unabsorbed flux\\
 & (s) &(0.3-10 keV) & \multicolumn{2}{c}{(erg cm$^{-2}$ s$^{-1}$, 2-10 keV)} \\
2005 May 15 & 62.6 & 0.094 & 5.6$\times$10$^{-12}$ & 6.3$\times$10$^{-12}$ \\
2005 November 01 & 456.3 & 0.019 & 1.1$\times$10$^{-12}$ &  1.3$\times$10$^{-12}$ \\
2006 February 07 & 138.6 & 0.043 & 2.5$\times$10$^{-12}$ & 2.9$\times$10$^{-12}$ \\
2007 February 02 & 193.1 & 0.038 & 2.3$\times$10$^{-12}$ & 2.5$\times$10$^{-12}$ \\
2008 February 21 & 113.7 & 0.067 & 4.0$\times$10$^{-12}$ & 4.5$\times$10$^{-12}$ \\
combined dataset & 964.3 & 0.010 & 5.9$\times$10$^{-13}$ & 6.7$\times$10$^{-13}$ \\
\end{tabular}
\label{tab:xraylog}
\end{center}
\end{table}

\clearpage

\clearpage

\clearpage

\end{document}